\documentclass[useAMS,usenatbib,referee]{mn2e}
\usepackage{graphicx}
   \title{White Dwarf Mass Distribution in the SDSS}

   \author[Kepler et al.]{S. O. Kepler$^{1}$\thanks{
kepler@if.ufrgs.br},
		S. J. Kleinman$^{2}$,
		A. Nitta$^{3}$,
		D. Koester$^{4}$,
		B. G. Castanheira$^{1}$,
\newauthor
		O. Giovannini$^{5}$,
		A. F. M. Costa$^{1}$,
          and
		L. Althaus$^{6}$\\
$^{1}$Instituto de F\'{\i}sica, Universidade Federal do Rio Grande do Sul,
              91501-900  Porto-Alegre, RS, Brazil\\
$^{2}$Subaru Telescope, 650 N. A'Ohoku Place, Hilo Hawaii, 96720, USA\\
$^{3}$Gemini Observatory, Hilo, Hawaii, 96720, USA\\
$^{4}$Institut f\"ur Theoretische Physik und Astrophysik, Universit\"at Kiel,
  24098 Kiel, Germany\\
$^{5}$ Universidade de Caxias do Sul, 95070-560 Caxias do Sul, RS, Brazil\\
$^{6}$ Facultad de Ciencias Astron\'omicas y Geof\'{\i}sicas, Paseo del Bosque S/N, 
(1900) La Plata - Argentina
}
\begin{document}
\date{Accepted 2006 December 6. Received 2006 November 22; in original form 2006 August 1}

\pagerange{\pageref{firstpage}--\pageref{lastpage}} \pubyear{2006}
   \maketitle

\label{firstpage}
  \begin{abstract}
We determined masses for the 7167 DA and 507 DB white dwarf stars classified as single
and non-magnetic
in data release four of the Sloan Digital Sky Survey (SDSS). 
We obtained revised $T_{\mathrm{eff}}$ and $\log g$ determinations for the most
massive stars by fitting 
the SDSS optical spectra with a synthetic spectra grid derived from model
atmospheres extending to $\log{g}=10.0$. We also calculate radii from
evolutionary models and create volume-corrected mass distributions for our
DA and DB samples.
The mean mass for the DA stars brighter than g=19 and hotter than 
$T_{\mathrm{eff}}=12\,000$~K is
$\langle{\cal{M}}\rangle_\mathrm{DA}\simeq 0.593\pm 0.016~{\cal{M}}_\odot$.
For the 150 DBs brighter than g=19 and hotter than $T_{\mathrm{eff}}=16\,000$~K, 
we find $\langle {\cal{M}}\rangle_\mathrm{DB}=0.711\pm 0.009 {\cal{M}}_\odot$.
It appears the mean mass for DB white dwarf stars may be significantly larger than that
for DAs.
We also report the highest mass white dwarf stars ever found, up to  $1.33~{\cal{M}}_\odot$.
\end{abstract}


\section{Introduction}

White dwarf stars are the end product of evolution of all stars
with initial masses
up to around $9~{\cal{M}}_\odot$ and their distribution
contains information about star formation history and
subsequent evolution in our Galaxy.
As
the most common endpoints of stellar evolution, white dwarf stars account for around 97\%
of all evolved stars. 
Considering  there has not yet been enough time for any
of them to cool down to undetectability,
they can also provide independent information about the age of
the Galaxy. 
Through an
initial--final mass relation (IFMR), we can also study mass
loss throughout the stellar evolution process.
Because white dwarf progenitors
lose carbon, nitrogen, and oxygen
at the top of the asymptotic giant branch, they are significant
contributors to the chemical evolution of the Galaxy and
possibly an important source of life
sustaining chemicals.

\citet{scot} published the spectra of 2551 
white dwarf stars
in the Sloan Digital Sky Survey (SDSS) Data Release 1 (DR1), covering 
$1360~{\mathrm{deg}}^2$.
\citet{scot06} extended the white dwarf spectroscopic identifications
to the 4th SDSS Data Release with a total of 9316 white dwarf stars reported,
more than doubling the number of spectroscopically identified stars 
\citep{McS}. 
In both works, the authors fit the entire optical
spectra from 3900\AA\ to 6800\AA\ to DA and DB grids 
of synthetic spectra derived from model atmospheres
calculated by Detlev Koester, up to $\log g=9.0$. 
Their fits include SDSS imaging photometry and allow for 
refluxing of the models by a low-order polynomial to
incorporate effects of unknown reddening and spectrophotometric errors.
The SDSS spectra have a mean g-band signal--to--noise ratio SNR(g)$\approx 13$ for all DAs, 
and SNR(g)$\approx 21$ for those brighter than g=19.

This large sample of stars with spectroscopic fits gives us a new
opportunity to fully explore the white dwarf mass distribution.
Understanding the white dwarf mass distribution offers insights into mass
loss during stellar evolution, the IFMR, and has bearings on close binary
star evolution.  Our report, as well as many previous studies,
detect a substantial fraction of low mass white dwarf stars
that theoretically cannot have evolved as single stars, because the age of the
Universe is smaller than their presumed lifetimes on the main sequence.

\citet{scot} notice an increase in mean $\log g$ for stars cooler than
$T_{\mathrm{eff}}=12000$~K, but  caution the trend might not be real, indicating
a problem in the data or fit technique, instead.  The trend has persisted
into the larger catalog of \citet{scot06}.
\citet{madej} analyzed the \citet{scot} sample of fits and calculated the
corresponding SDSS DR1 pure hydrogen atmosphere (DA) 
white dwarf mass distribution.
As expected from the $\log g$ trend, they found that 
the mean mass also increased below $T_{\mathrm{eff}}=12000$~K.
Their Table~1 presents all previous mean mass determinations,
producing an average of $0.57~{\cal{M}}_\odot$, and a most populated peak
at $0.562~{\cal{M}}_\odot$ for the 1175 stars hotter than $T_{\mathrm{eff}}=12000$~K.
They did not study the potentially highest mass stars with $\log g>9$, because they were
limited by the  stellar atmosphere fit by \citet{scot} which 
artificially pegged stars near the upper $\log g=9.0$ boundary to
the boundary itself.

The increase in mean masses fitted from optical spectra
below $T_{\mathrm{eff}}=12000$~K has been
seen prior to \citet{scot} and has been discussed since
\citet{Bergeron91} and \citet{koester91}. It  is usually dismissed as due to
problems in the models: either convection bringing up subsurface
He to the atmosphere, increasing the local pressure, or 
problems with the treatment of the
hydrogen level occupation probability. The new larger SDSS data set,
however, now gives another opportunity to explore this trend and evaluate
its cause.

Most reported white dwarf mass determinations have been derived by comparing the optical spectra
with model atmospheres, as with \citet{scot} and \citet{scot06}.
For the DA stars,
the H7, H8 and H9 lines, in the violet, are the most sensitive
to surface gravity because they are produced by electrons at
higher energy levels, those most affected by neighboring atoms. However, these
lines are also in the region where the atmospheric extinction
is the largest and typical CCD detectors are the least sensitive.
As a consequence, most studies used only the line profiles in their fits,
avoiding the dependence on often uncertain flux calibrations.
The SDSS white dwarf spectra have good flux calibration and acceptable
SNR redwards of 4000\AA.
The published SDSS catalog therefore fits the {\it entire} optical spectrum, and
not just the H lines, as has been traditionally done.  The rationale for
this approach is the good, uniform spectrophotometry and corresponding
broad band photometry that can be used in the fits.  In addition, a
low-level re-fluxing is allowed to take out large errors in
spectrophotometry and any unknown reddening effects.

In this paper, we will compare the measured white dwarf mass distributions from
\citet{scot} and \citet{scot06} with previous determinations and attempt to
assess the reason for the observed increase in mass for lower
temperatures.  We will also explore the observed mean masses and analyze
the two different fitting techniques: line profile vs. whole spectrum, to see
the effects on the resulting mass distributions.

\section{Data and Models}
The SDSS imaged the high Galactic latitude sky in five passbands:
u, g, r, i and z, and obtained spectra from 3800\AA\ to 9200\AA\ 
with a resolution of $\approx$ 1800 using a twin fiber-fed spectrograph
\citep{York}.
Since we are primarily interested in the mass distribution here, we
selected 
only
the single DA and DB stars with 
$\log g-\sigma_{\log g}\geq 8.5$ 
and 
$\log g+\sigma_{\log g}\leq 6.5$ 
from the \citet{scot06} sample and refit them with an
expanded grid of models (see below), using the same {\it autofit} routine as in
\citet{scot06} and thoroughly described in \citet{scot}.
We excluded all stars 
classified by \citet{scot06} as having either
a detectable magnetic field or a companion, metal lines, DABs, and  DBAs.

Our model grid 
\citep{Finley,koester2001}
is similar 
to that used by \citet{scot06},
but extended in $T_{\mathrm{eff}}$ 
and $\log g$ (100\,000~K $\leq T_{\mathrm{eff}} \leq$ 6000~K,
$10.0\leq \log g \leq 5.0$)
and denser.
We chose the ML2/$\alpha=0.6$ parameterization for convection as
demonstrated by \citet{bergeron95} to give internal consistency between
temperatures derived in the optical and the ultraviolet, photometry,
parallax, and with gravitational redshift. 
ML2 corresponds to the
\citet{bohm} description of the mixing length theory and
$\alpha=\ell/\lambda_P$ is the ratio of the mixing length to the 
pressure scale height.
The models include the $H_2^+$ and $H_2$ quasi-molecular opacities and
only Stark \citep{Lemke}  and Doppler broadening, so 
the line profiles are not precise for
$T_{\mathrm{eff}} < 8500$~K.

Even though \citet{napiw} and \citet{liebert} discuss the
necessity of using NLTE atmospheres for the stars
hotter than 40\,000~K, all quoted values are from LTE models,
as they also show the NLTE corrections are not dominant,
and our number of hot stars is small.

To calculate the mass of each star
from the $T_{\mathrm{eff}}$ and $\log g$ values
obtained from our fits, we used the evolutionary models of \citet{Wood} and
\citet{Althaus} with C/O cores  up to
$\log g=9.0$, and O/Ne cores for higher gravities,
$M_{\mathrm{He}}=10^{-2}\,M_*$, and $M_{\mathrm{H}}=10^{-4}\,M_*$, 
or $M_{\mathrm{H}}=0$, to estimate stellar radii for DAs and DBs,
respectively. The radius is necessary to convert surface gravity to mass.

\section{Analysis}

Before exploring the mass distributions, we wanted to examine the
different fitting techniques used in the available data sets --- the
traditional line profile technique and the SDSS whole spectrum approach.
We therefore simulated spectra with differing SNRs by adding random
noise to our models and fit them with our own set of both line profile
and whole spectrum fitting routines.
Our Monte Carlo simulations show that in the low SNR regime, SNR$\leq 50$, fitting the
whole spectra and not just the line profiles gives {\it more} accurate
atmospheric parameters,
as long as the flux calibration
or interstellar reddening uncertainties do not dominate.
We estimate an uncertainty of around
$\Delta T_{\mathrm{eff}}\simeq 500$~K and 
$\Delta \log g \simeq 0.10$ at SNR=40
for the whole spectra
fitting. 
For SNR=20, similar to the average SDSS spectra for $g<19$, 
our simulations indicate
$\Delta T_{\mathrm{eff}}\simeq 750$~K and 
$\Delta \log g \simeq 0.15$. We do not report in
this paper on the
mass distribution for the stars fainter than g=19 because their
smaller SNR lead to large uncertainties.
Our simulations did not indicate systematic trends between the two approaches.  

Although \citet{scot} and \citet{scot06} compared their fits' internal
errors by fitting
duplicate spectra, they did not display their results as a function of
temperature.  \cite{Soar} specifically analyzed 
109 duplicate spectra SDSS DAs
with 13000~K $\geq T_{\mathrm{eff}} \geq$ 10000~K, near the region where
the fit $\log{g}$s start to increase.
They showed that the
mean fit differences were
$\sigma_{T_{\mathrm{eff}}}\simeq 300$~K and
$\sigma_{\log g}\simeq 0.21$~dex for the same object but
different observations. These values  are
larger than the internal uncertainty of the fits,
but in general within $3\sigma$ of each other,
as in \citet{scot} and \citet{scot06}. 
We thus conclude that the uncertainties in \citet{scot06} are reasonable
and can now
analyze the results without attributing any noted irregularities to the
fitting process.

\citet{gemini}, however, 
compare SDSS spectra with new SNR(g)$\simeq 100$ spectra acquired with GMOS on
the Gemini 8~m telescope for four white dwarf stars around
$T_{\mathrm{eff}}\simeq 12\,000$~K. Their fits suggest
that published SDSS optical spectra fits
overestimate the mass by
$\Delta {\cal{M}}\simeq 0.13~{\cal{M}}_\odot$,
because of the correlation between the
derived $T_{\mathrm{eff}}$ and $\log g$ --- a small increase in 
$T_{\mathrm{eff}}$ can be compensated by a small decrease in
$\log g$.  Our simulations
indicate
this discrepancy
is concentrated only in the region around the Balmer line maximum,
14000~K$\geq T_{\mathrm{eff}} \geq$11000~K.

%

\begin{figure}
   \centering
   \includegraphics[width=0.5\textwidth]{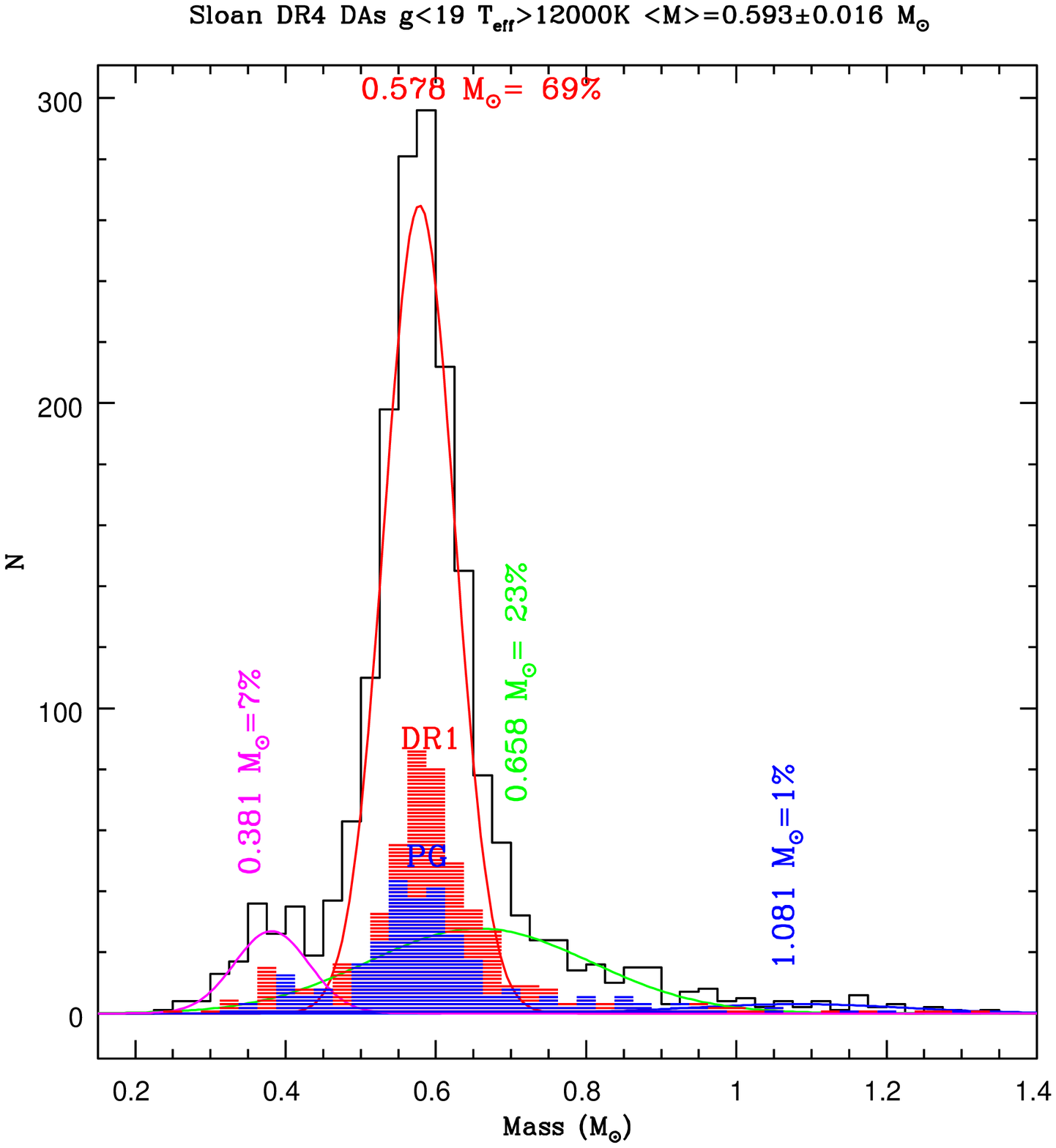}
   \includegraphics[width=0.5\textwidth]{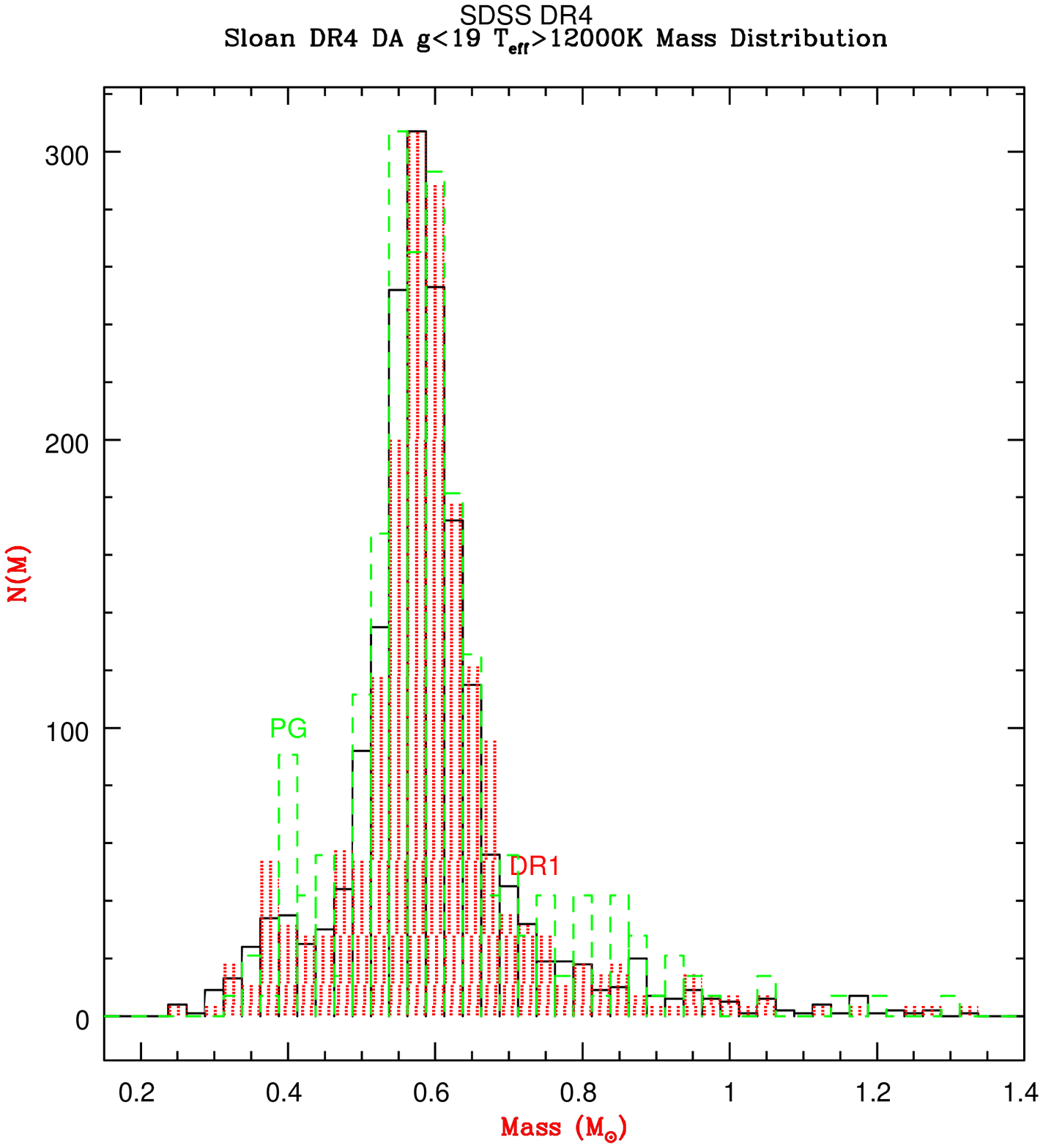}
      \caption{Histogram for the 1733 DA stars
 brighter than g=19 and hotter than
$T_{\mathrm{eff}}=12\,000$~K, compared to the PG survey
published by \citet{liebert} and the SDSS DR1 sample published 
by \citet{madej}.
Gaussian fits detailed in Table~\ref{thist1912} are also shown.
Our bins are $0.025~{\cal{M}}_\odot$ wide.
The second graph shows the DR1 and PG survey data normalized to the DR4
sample, even though those samples are smaller and therefore have 
significantly larger errorbars.
              }
         \label{hist}
   \end{figure}

To explore the increasing mass trend in more detail, we restricted our
sample
to the 1733 stars both brighter than g=19 and hotter than 
$T_{\mathrm{eff}}=12\,000$~K and obtained an average DA mass of
$\langle{\cal{M}}\rangle_\mathrm{DA}=0.593\pm 0.016~{\cal{M}}_\odot$. 
The distribution for this hot and bright sample,
shown in Fig.~\ref{hist},
is similar
to that of the Palomar Green survey published by \citet{liebert}.
They studied a complete sample of 348 DA stars
with SNR$\geq$ 60 spectra and determined
atmospheric parameters by spectral fitting via
the line profile fitting technique,
using models up to $\log g=9.5$. 
They found a peak in the mass histogram  
at $0.565~{\cal{M}}_\odot$ containing 75\% of the sample, 
a low mass peak with
$0.403 ~{\cal{M}}_\odot$ containing 10\% of the sample, and
a high mass peak at
$0.780~{\cal{M}}_\odot$ containing 15\% of the stars.
They fit their  mass histogram (PG mass histogram from hereafter)
with three Gaussian profiles:
$0.565~{\cal{M}}_\odot$ with $\sigma\simeq 0.080~{\cal{M}}_\odot$,
$0.403~{\cal{M}}_\odot$ with $\sigma\simeq 0.023~{\cal{M}}_\odot$,
and a broad high-mass component at
$0.780~{\cal{M}}_\odot$ with $\sigma\simeq 0.108~{\cal{M}}_\odot$.
They found more stars above  $1~{\cal{M}}_\odot$
than can be described by the three Gaussians they fit.
\citet{vennes97}, \citet{vennes99}, and
\citet{marsh} also find an excess of white dwarf stars
with masses above  $1~{\cal{M}}_\odot$ in their sample of $T_{\mathrm{eff}}>23\,000$~K white dwarf stars.

The overall mass distribution of our bright sample matches well with that
of the previous standard PG survey sample.  We now explore the distribution
with temperature.

\begin{figure}
   \centering
   \includegraphics[width=\textwidth]{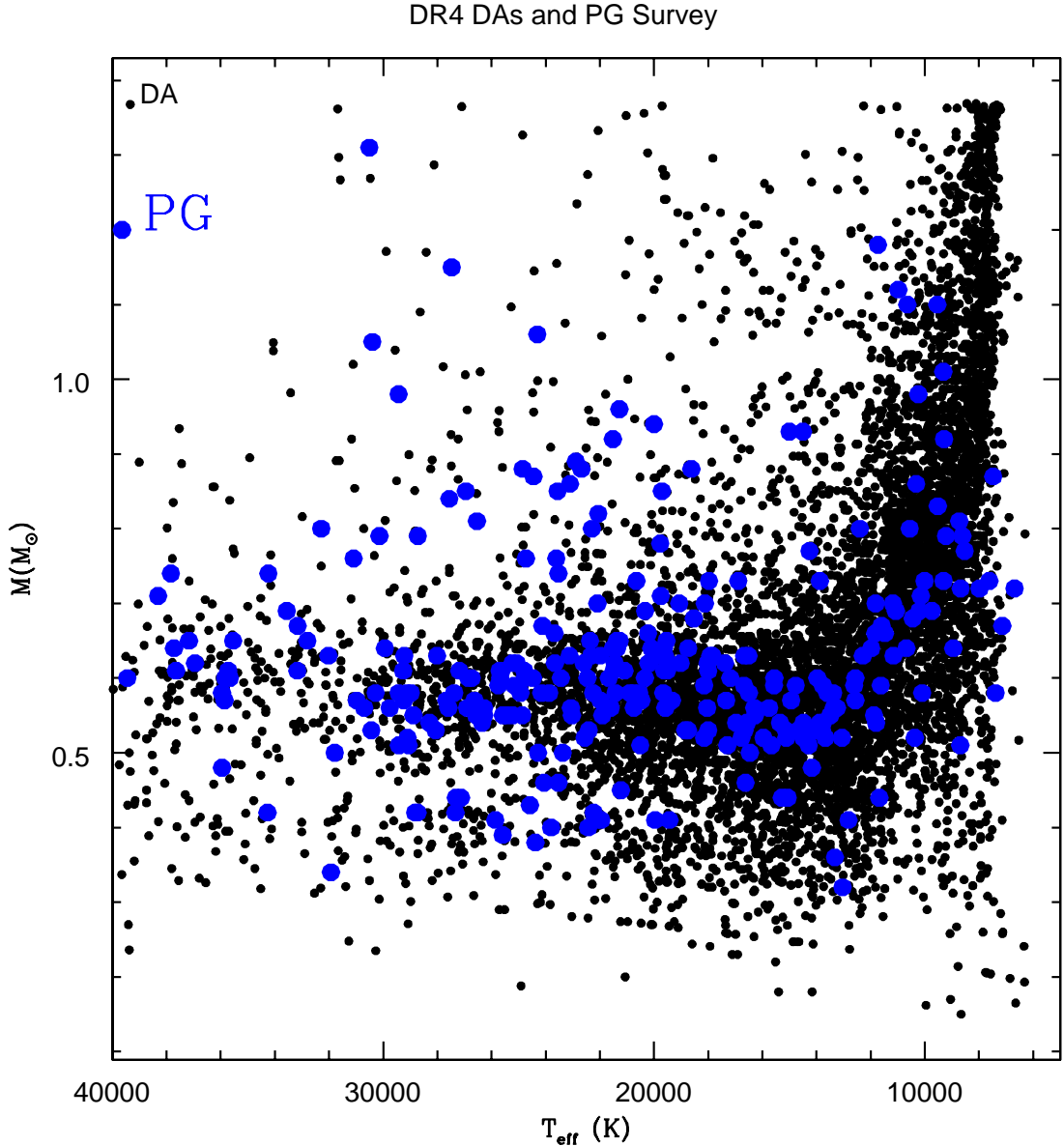}
      \caption{Masses for all 3595 DA white dwarf stars
 brighter than g=19 and cooler than 40\,000~K, showing an increase in mean
mass for lower $T_{\mathrm{eff}}$. The large solid (blue) circles are the
values published by \citet{liebert}, showing the increase in mass at lower
$T_{\mathrm{eff}}$ is also present in their sample, which uses a totally
independent grid of models and fitting technique.
              }
         \label{teff19}
   \end{figure}

In Fig.~\ref{teff19}, we show the
mass distribution vs. temperature for DA stars brighter than
$g=19$ along with the similar distribution from \citet{liebert}.
Again, we see the distributions are roughly equivalent and we see an
increase in measured mass at lower temperatures.
Our histograms use 0.025~${\cal{M}}_\odot$ bins 
(N/dm=constant)
because that is the approximately
mean uncertainty in our mass determinations.

\begin{figure}
   \centering
   \includegraphics[width=\textwidth]{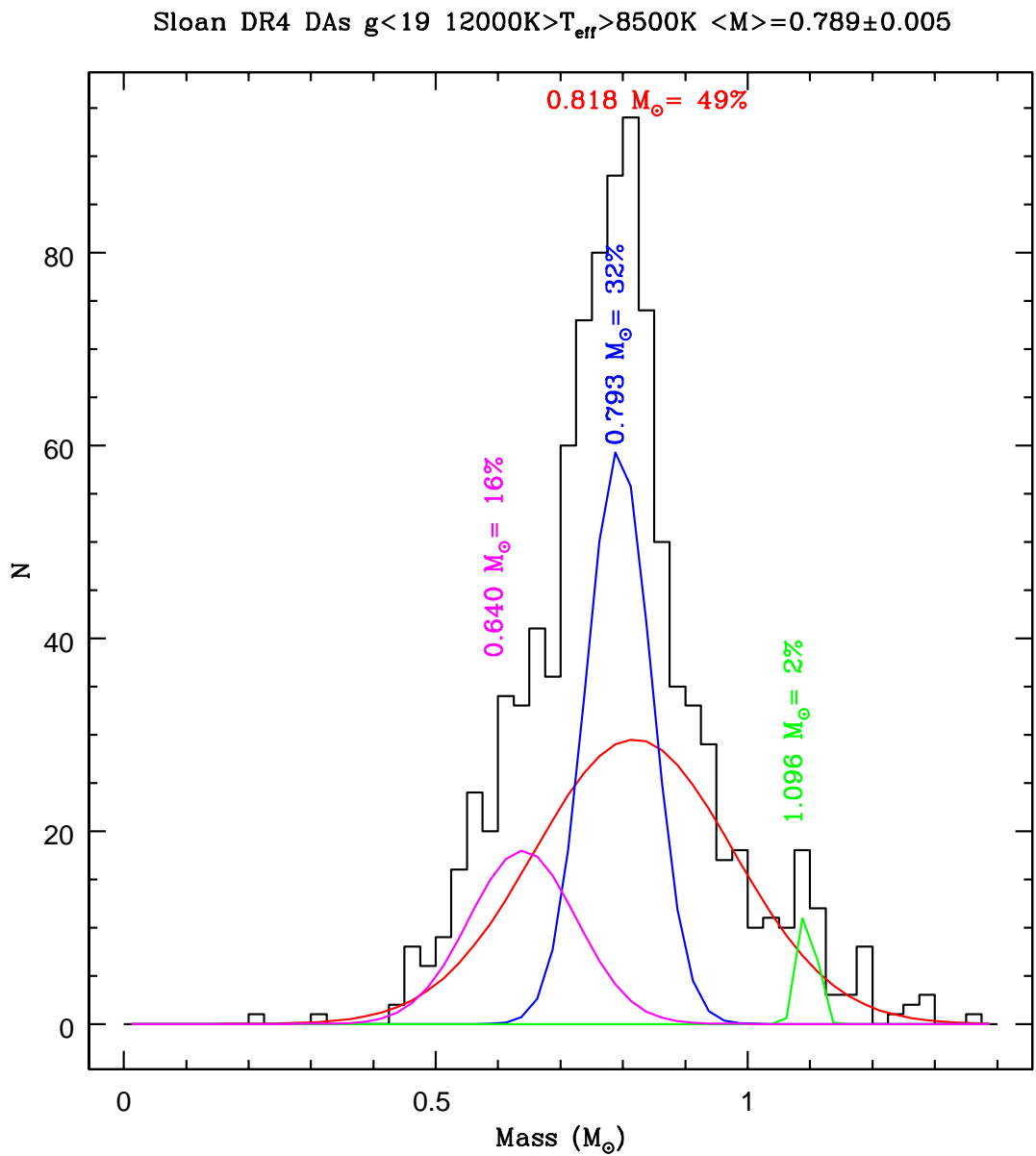}
      \caption{Histogram for the 964 DA stars
 brighter than g=19,
with $12\,000~\mathrm{K} \geq T_{\mathrm{eff}} \geq 8\,500$~K
along with the fit Gaussians as detailed in Table~\ref{thist191285}.
         \label{hist191285}
              }
   \end{figure}

To explore the region of increasing mass further,
Fig.~\ref{hist191285}
shows the mass histogram only for the 964 DAs 
brighter than g=19,
and $12\,000~\mathrm{K} \geq T_{\mathrm{eff}} \geq 8\,500$~K,
for which we obtain
$\langle{\cal{M}}\rangle_\mathrm{DA}^\mathrm{cool}\simeq 0.789\pm 0.005~{\cal{M}}_\odot$.
We have excluded the stars cooler than $T_{\mathrm{eff}}=8500$~K from
our mass histograms because our cooler atmospheric models are not accurate for
$\log g$ determination, as explained earlier.

Tables~\ref{thist1912}~and~\ref{thist191285} detail the Gaussian fits we
made for the histograms of Figures~\ref{hist}~and~\ref{hist191285}
respectively, with
\begin{equation}
N = \sum_i a_i  \exp\left[-\frac{({\cal{M}}-{\cal{M}}_i)^2}
{2\sigma_i^2}\right]
\label{eqN}
\end{equation}

\begin{table}
\centering
\caption{Gaussian fits for the  $T_{\mathrm{eff}}\geq 12\,000$~K
and $g\leq 19$ histogram, seen in Fig.~\ref{hist}.
\label{thist1912}}
\begin{tabular}{crccr}
\hline
i&$a_i$&$M_i ({\cal{M}}_\odot)$&$\sigma_i$&Fraction\\
\hline
1&264.8 &0.578 &0.047&69\%\\
2& 27.8 &0.658 &0.149&23\%\\
3& 27.0 &0.381 &0.050&7\%\\
4&  3.0 &1.081 &0.143&1\%\\
\hline
\end{tabular}
\end{table}

\begin{table}
\centering
\caption{Gaussian fits for the 964 DAs with  $12\,000~\mathrm{K} \geq T_{\mathrm{eff}}\geq 8\,500$~K and $g\leq 19$ histogram.
\label{thist191285}}
\begin{tabular}{crccr}
\hline
i&$a_i$&$M_i ({\cal{M}}_\odot)$&$\sigma_i$&Fraction\\
\hline
1& 29.5 &0.818 &0.160 &49\%\\
2& 59.6 &0.793 &0.052 &33\%\\
3& 18.0 &0.640 &0.086 &16\%\\
4& 13.4 &1.096 &0.136 &2\%\\
\hline
\end{tabular}
\end{table}

Figure~11 of \citet{liebert} also shows an increase in mass below
$T_{\mathrm{eff}}=12\,000$~K (see Fig.~\ref{teff19}),
even though they have a limited 
number of cooler stars due to color selection effects in the PG survey.
It is important to note that the model atmospheres used in
\citet{liebert} and the line profile fitting technique they use
are totally independent of our own.  Therefore,
if the observed increase in the measured gravity
with temperature is merely an artifact of the models,
then similar problems must be present in two independent
groups of models and fitting techniques.  We are thus gathering increasing
evidence that either a) both DAs and DBs really do have higher mean masses
at lower temperatures, or b) there is a common artifact in the model used
for all white dwarf spectral fitting.

%
%
\section{Why we do not trust masses for 
$T_{\mathrm{eff}}<12000$~K for DA STARS}
\label{discus}
\citet{bergeron95} measured an increase in the mean mass
for the ZZ Ceti star sample around 13000~K$>T_{\mathrm{eff}}>$11000~K,
but indicated it might come from a selection effect 
because the
discovery of pulsating stars might have preferred higher mass stars.
\citet{arras}, e.g., show that the 
white dwarf
pulsators with lower masses should pulsate
at cooler temperatures. Our sample of 351 bright stars
in the same temperature range
show a similar increase in mass compared to the hotter sample, 
but we have not been biased by the
pulsators, so an observational bias is not the cause for the
increase in mass detected.

The simple expectation that massive stars cool faster than
their less massive counterparts does not hold for 
$T_{\mathrm{eff}}\leq 10 \,000$~K, as the most
massive stars have smaller radius and, therefore, their cooling
slows down after a few e-folding timescales.
Another possible explanation for an increased mean mass at lower 
effective temperatures is the presence of otherwise undetected He 
at the surface,
broadening the observed H lines and thus mimicking a higher $\log g$.
Theoretical models (e.g.  \citet{fw91})
indicate that only for DAs with hydrogen layer masses
below ${\cal{M}}_H=10^{-10} {\cal{M}}_*$ will He mix around $T_{\mathrm{eff}}=10\,000$~K
and, if ${\cal{M}}_H=10^{-7} {\cal{M}}_*$, only below
$T_{\mathrm{eff}}\simeq 6\,500$~K.
However, all seismologically measured H
layer masses are ${\cal{M}}_H>10^{-7} {\cal{M}}_*$
\citep{Bradley06, Bradley01, Bradley98, Fontaine}.
Since our increased mass trend happens significantly
hotter than $T_{\mathrm{eff}}=6\,500$~K, He contamination cannot account for the observed increase
in mass at lower temperatures, unless the more distant stars 
studied here have significantly
thinner H layers. \citet{Lawlor} models show around 3\% of the DAs
could have ${\cal{M}}_H\sim 10^{-9} {\cal{M}}_*$, but not thinner.
Therefore, there are not enough stars with thin H layers,
at any rate, to account for our excess of
massive objects.

\citet{leeann} proposes a possible physical model for increasing WD masses
at lower temperatures.  She suggests that low metallicity AGBs will produce higher
mass white dwarf stars, probably around $1{\cal{M}}_\odot$,
because the relatively lower mass loss 
expected for low metallicity AGB stars 
increases the mass of the core prior
to 
the star moving out of the AGB.  Since the earlier generations of white
dwarf stars which have now cooled more than their later cohorts,  presumably
came from lower-metallicity progenitors, this mechanism could explain a mass
increase at lower white dwarf temperatures.
If  we extend this concept to globular clusters though, we would expect
the mass of the white dwarfs in globular clusters to be larger than
the mean mass of our stars cooler than 10\,000~K, 
which is not observed \citep{Moehler, Richer}.  So again, we are left with a
discarded explanation of the observed mass increase.

\begin{figure}
   \centering
   \includegraphics[width=\textwidth]{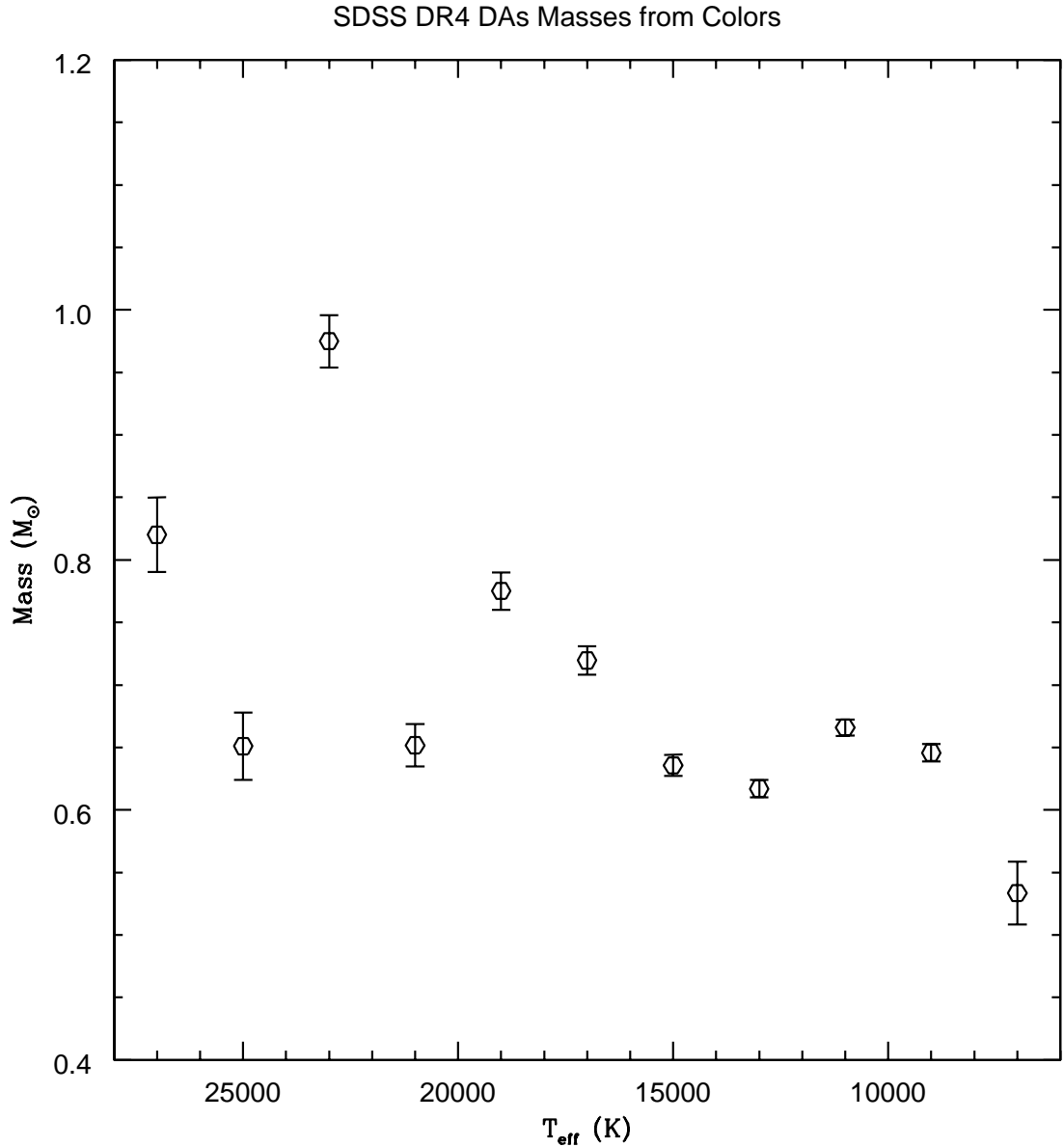}
      \caption{Masses for all 7167 DA white dwarf stars
derived comparing only the SDSS colors (u-g) and (g-r) with those
predicted from the atmospheric models. For
$T_{\mathrm{eff}}\geq 20\,000$K
and $T_{\mathrm{eff}}\leq 9\,000$~K,
the colors are degenerate in mass. 
              }
         \label{cor}
   \end{figure}

An interesting clue to the problem may be found in
\citet{Koester06}, which used SDSS photometry alone to make a mass
estimate.  Their 
cool white dwarf stars show mean masses similar to those of the hotter stars.
Our mass determinations using photometric colors only,
shown in Fig.~\ref{cor}, is
derived comparing only the SDSS colors (u-g) and (g-r) with those
predicted from the atmospheric models convolved with SDSS filters. They
do not show any
increase in mass with decreasing $T_{\mathrm{eff}}$. Because of their
larger uncertainties than the spectra fitting, we binned the results
in 2000~K bins.
This result suggests that any problem in the models is 
mainly restricted to the line profiles, not the
continua, which dominate the broadband photometric colors.

Thus, we are mainly left with the possibility raised by
\citet{koester91} that an increase in
mass with lower temperatures could be due to the treatment of neutral particles
in model atmospheres with the Hummer-Mihalas formalism.
\citet{bergeron95}, however, suggests that the neutral particles
are only important below $T_{\mathrm{eff}}\simeq 8000$~K which is certainly
lower than where we see the trend begin.  It seems the only
remaining explanation is that accurate modeling of
neutral particles will indeed show an effect for DAs near 12\,000~K.


\begin{figure}
   \centering
   \includegraphics[width=\textwidth]{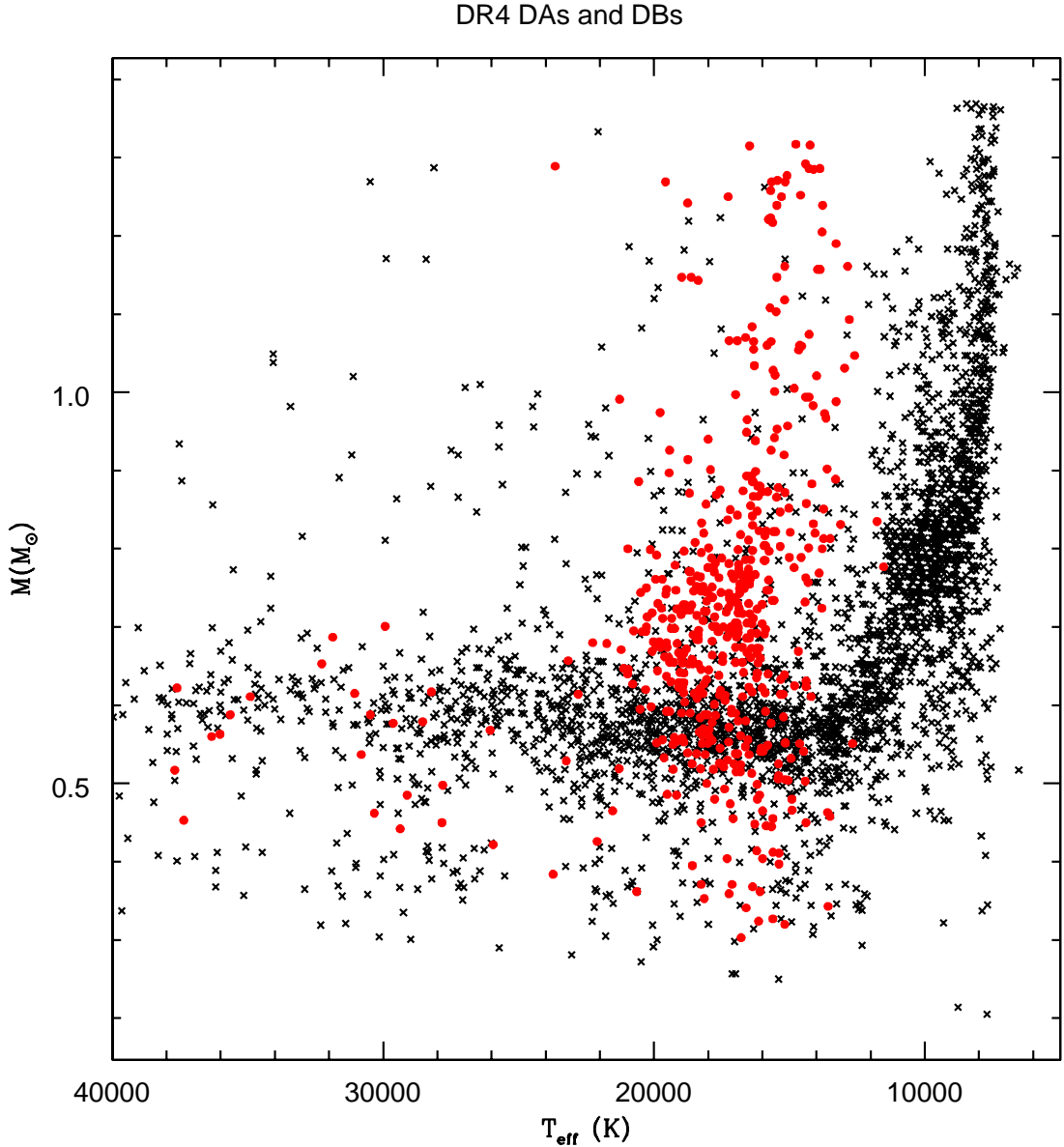}
      \caption{Masses for all 5718 DAs (crosses)
 cooler than 40\,000~K
and 507 DBs (filled circles)
showing the continuous increase
in average mass at lower $T_{\mathrm{eff}}$. 
              }
         \label{teff}
   \end{figure}
\section{DB WHITE DWARFS}
We determined masses for the \citet{scot06} DBs from their fit temperatures
and gravities using evolutionary grids
of \citet{Althaus,Althaus01}.
The Althaus models
use time resolved diffusion throughout evolution.
\citet{Travis05} and \citet{Travis05a} discuss asteroseismological results
in DBs, showing the observations are consistent with the layer masses
predicted by current diffusion theory.
Fig.~\ref{teff} shows we find an increase in the measured
surface gravity below $T_{\mathrm{eff}} \simeq 12\,000$~K for
DAs and a similar increase below $T_{\mathrm{eff}} \simeq 16\,000$~K for DBs. 
For the 208 DBs brighter than g=19, we find 
$\langle {\cal{M}}\rangle_\mathrm{DB}^\mathrm{all}=0.785\pm 0.013 {\cal{M}}_\odot$
For the 150 DBs brighter than g=19 and hotter than $T_{\mathrm{eff}}=16\,000$~K, we find
$\langle {\cal{M}}\rangle_\mathrm{DB}=0.711\pm 0.009 {\cal{M}}_\odot$. 
Both measurements are considerably
larger than the $0.593\,{\cal{M}}_\odot$ mean mass value for the bright and hot DA sample.
A similar larger (relative to that of the DAs) DB mean mass value has been
previously reported by \citet{koester2001} who
obtained a
$\langle  {\cal{M}} \rangle_\mathrm{DB}=0.77$ for the
18 DBs they observed with UVES/VLT, including stars down to 
$T_{\mathrm{eff}}\sim 16\,000$~K.
Others, however, find lower mean DB masses, more similar to those of the
DAs.
\citet{Oke} derived $\langle  {\cal{M}} \rangle_\mathrm{DB}=0.55\pm 0.10$ 
from their sample of
25 DBs ranging $30\,000~\mathrm{K} > T_{\mathrm{eff}} > 12\,000$~K, 
while \citet{Beauchamp} 
found $\langle  M \rangle_\mathrm{DB} = 0.59 \pm 0.01 {\cal{M}}_\odot$ 
for his 46 DBs, 
ranging $12\,000~\mathrm{K} \geq T_{\mathrm{eff}} \geq 31\,000$~K.
For the 34 DBs in \citet{barbaradb}, 
ranging $27\,000~\mathrm{K} \geq T_{\mathrm{eff}} \geq 13\,000$~K,
the mean is $\langle  {\cal{M}} \rangle_\mathrm{DB}=0.544\pm 0.05  {\cal{M}}_\odot$.


\begin{figure}
   \centering
   \includegraphics[width=\textwidth]{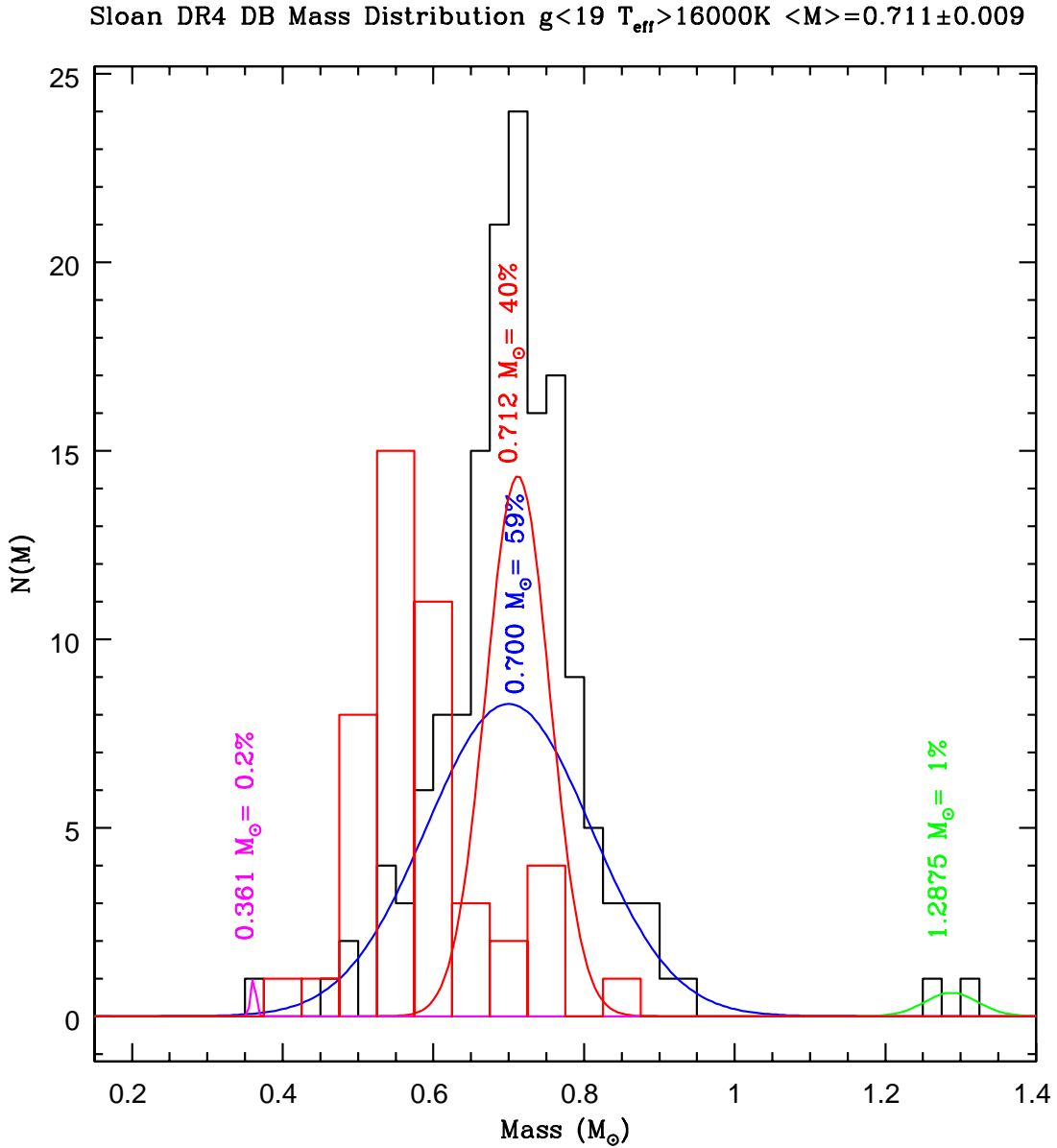}
      \caption{Histogram for the 150 DBs brighter than g=19 
and hotter than
$T_{\mathrm{eff}}=16\,000$~K in DR4.
         \label{histdb}
              }
   \end{figure}

The Gaussian fits 
for the 150 DBs brighter than g=19 and hotter than
$T_{\mathrm{eff}}=16\,000$~K are listed in Table~\ref{thistdb1916}.
The mass histogram for DBs is shown in Fig.~\ref{histdb}.
\begin{table}
\centering
\caption{Gaussian fits for the histogram of the 150 DBs brighter than g=19 and hotter than 
$T_{\mathrm{eff}}=16\,000$~K.
\label{thistdb1916}}
\begin{tabular}{crccr}
\hline
i&$a_i$&$M_i ({\cal{M}}_\odot)$&$\sigma_i$&Fraction\\
\hline
1&8.3 &0.700 & 0.109&59\%\\
2&14.3  &0.712 & 0.042&40\%\\
3&0.6  &1.288 & 0.035&1\%\\
\hline
\end{tabular}
\end{table}

\section{Observing Volume Correction}
In order to turn our observed mass distributions into a real analytical
tool, we must first correct the sample for completeness. We do this by the
$1/V_{\mathrm{max}}$ formalism.
$V_{\mathrm{max}}$ is the volume defined by the maximum
distance at which a given object would still appear
in a magnitude limited sample \citep{vmax}.
\citet{geijo} discuss white dwarf luminosity function completeness corrections and
conclude that for large samples,
the $1/V_{\mathrm{max}}$ method 
provides a reliable characterization of the white dwarf luminosity function.

\citet{liebert} find that 2.7\% of the stars in the PG sample have masses larger than
$1{\cal{M}}_\odot$ and, when corrected by $1/V_{\mathrm{max}}$,
22\% are above $0.8{\cal{M}}_\odot$.


We first calculated 
each star's absolute magnitude 
from the $T_{\mathrm{eff}}$ and $\log g$ values
obtained from our fits (for the extreme mass ones) or those of \citet{scot06}
(for the rest), convolving the synthetic spectra with the g filter 
transmission curve. We used the
evolutionary models of 
\citet{Wood} and
\citet{Althaus} with C/O cores  up to
$\log g=9.0$, and O/Ne cores for higher gravities, 
$M_{\mathrm{He}}=10^{-2}\,M_*$, and $M_{\mathrm{H}}=10^{-4}$ or $0\,M_*$, to
estimate stellar radii for DAs and DBs respectively. 
We do not claim that the SDSS spectroscopic sample is complete, but we do
contend that in terms of mass, there should be no preferential bias in the
target selection.
\citet{Harris} report that spectra are obtained for essentially
all white dwarf stars hotter than 22\,000~K.
Additional white dwarf stars down to 
$T_{\mathrm{eff}}=8000$~K are also found, but few cooler than that as
these stars overlap in color space with the F, G, and K main-sequence 
stars.
\citet{scot06} discuss the spectroscopic sample completeness,
which is around 60\% at $ 18<g<19.5$ for stars hotter than 
$T_{\mathrm{eff}}=12\,000$~K and around 40\% for cooler stars.
Our analysis is restricted to the sample brighter than g=19.

\begin{figure}
   \centering
   \includegraphics[width=\textwidth]{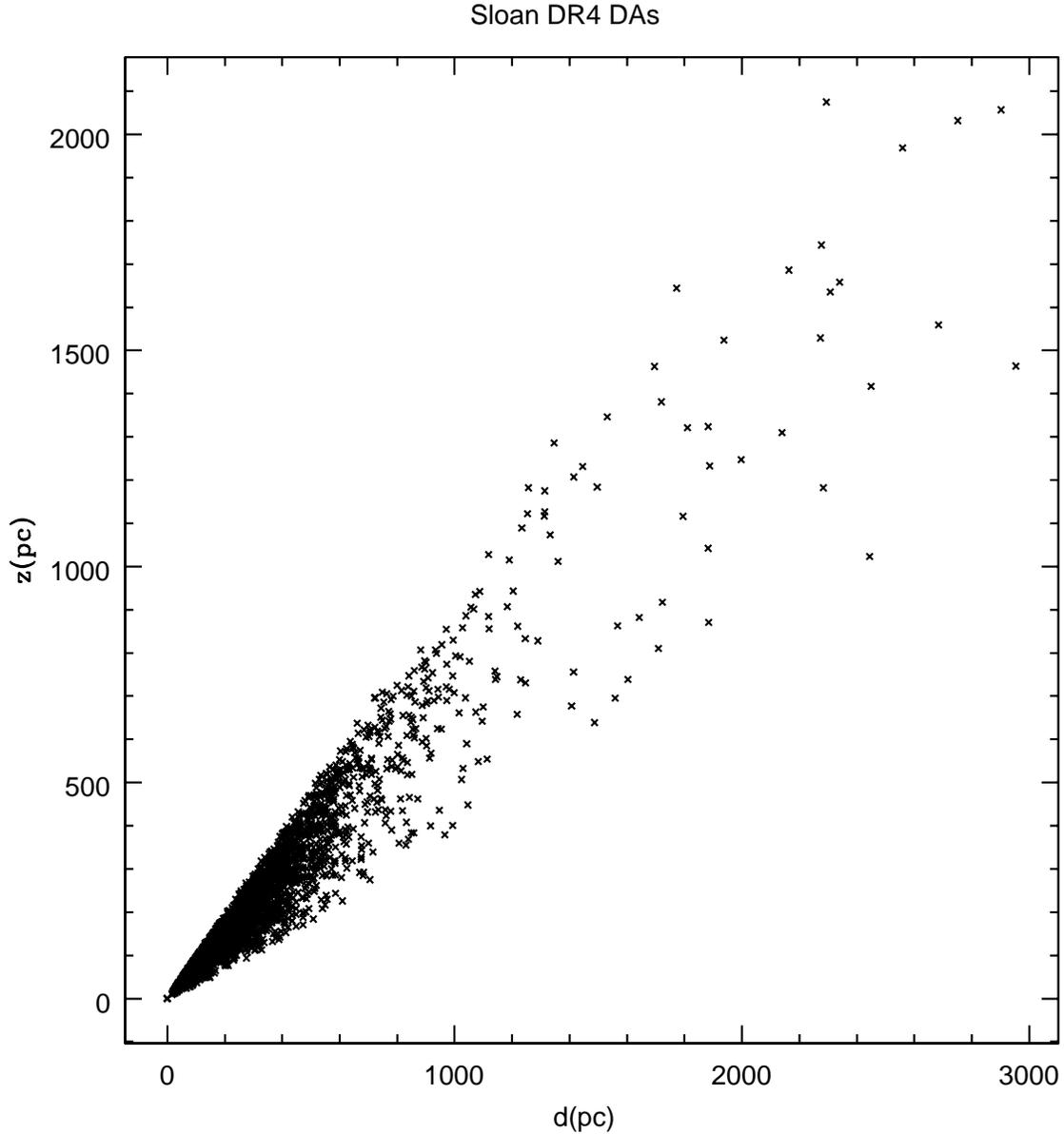}
      \caption{Distribution of distances, d, and height above
the Galactic plane, z, for DAs in the SDSS DR4.
              }
         \label{dist}
   \end{figure}

Once we had our calculated absolute magnitudes, we could estimate each
star's distance as shown in Fig.~\ref{dist}, neglecting any effects of
interstellar extinction.
The mean distance for our DA samples are:
$474\pm 5$~pc for the entire 7167 DA sample,
$302\pm 5$~pc for the stars brighter than g=19,
and $436 \pm 7$~pc for the stars brighter than g=19 and hotter than 
$T_{\mathrm{eff}}\simeq 12\,000$~K. 


\begin{figure}
   \centering
   \includegraphics[width=\textwidth]{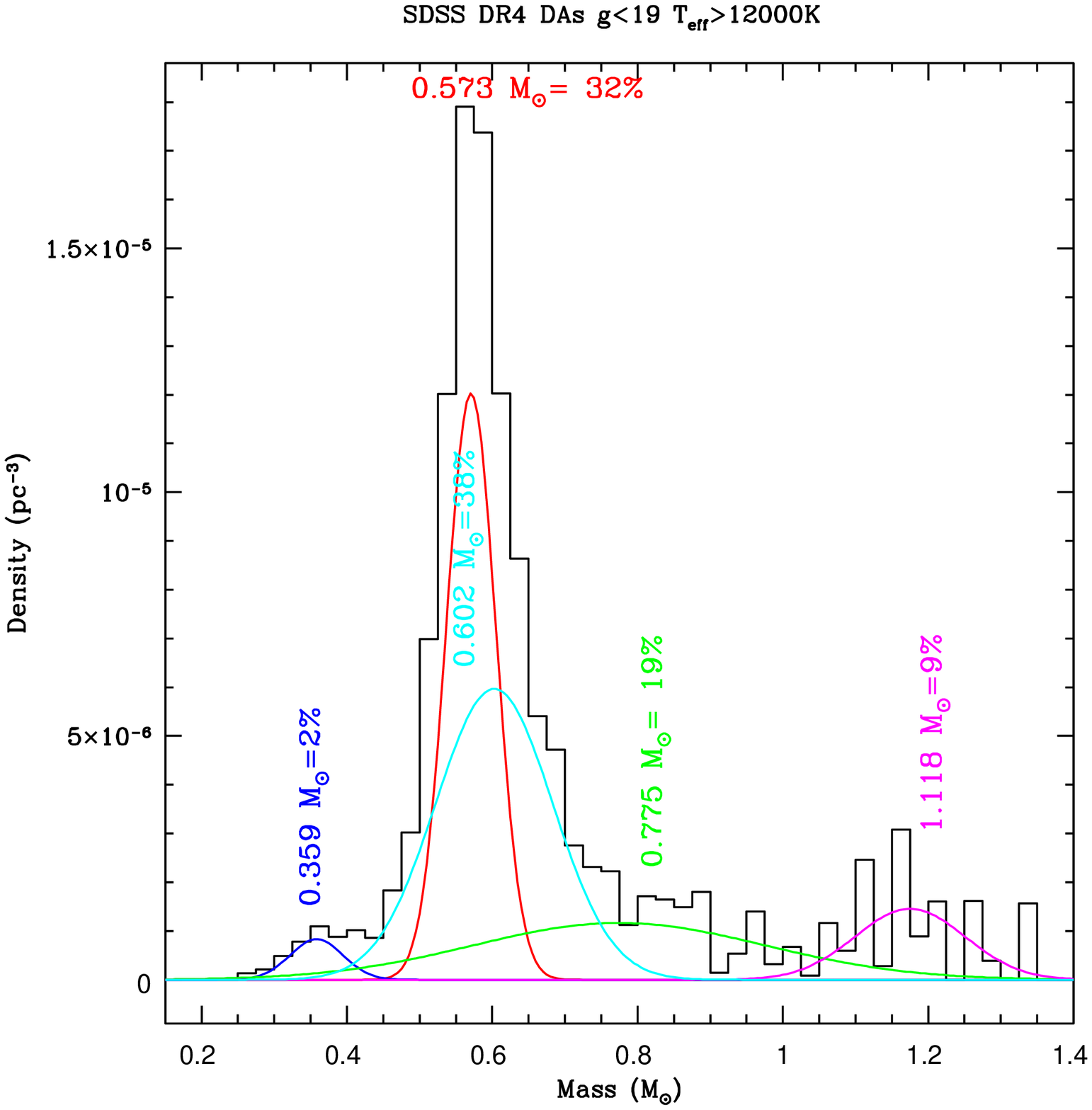}
      \caption{Histogram for the 1733 DA stars
 brighter than g=19 and hotter than
$T_{\mathrm{eff}}=12\,000$~K corrected by $1/V_{\mathrm{max}}$.
         \label{histv}}
   \end{figure}



For each star in our sample, we calculate
\[V_{\mathrm{max}} = \frac{4\pi}{3}\beta(r_{\mathrm{max}}^3-r_{\mathrm{min}}^3)
\exp(-z/z_0)\]
where $\beta$ is the fraction of the sky covered, 
0.1159 for the DR4 sample,
$r_{\mathrm{min}}$ is due to the bright magnitude limit,
g=15,
and $z_0$ is the disk scale height which we assume to be 250~pc,
as \citet{liebert} and \citet{Harris}, even though our height
distribution 
indicates $z_0\simeq 310$~pc.
\citet{vennes05} show that both the white dwarf stars in the 
SDSS DR1, and the 1934 DAs found in the 2dF 
($18.25 \leq b_J \leq 20.85$)
quasar surveys, belong to the thin disk of our Galaxy.
Using these data, they measured a scale height around 300~pc.
\citet{Harris} calculate the white dwarf luminosity function
from photometric measurements of the white dwarf stars
discovered in the SDSS survey up to DR3. They assume $\log g=8.0$ for
all stars and use the change in number per magnitude bin to
calculate the scale height of the disk, obtaining $340^{+100}_{-70}$~pc,
but adopt 250~pc for better comparison with other studies.
This volume includes the disk scale height as
discussed by \citet{Green}; \citet{Fleming}; and \citet{liebert}.
Each star contributes $1/V_{\mathrm{max}}$ to the local space
density. 

\begin{figure}
   \centering
   \includegraphics[width=\textwidth]{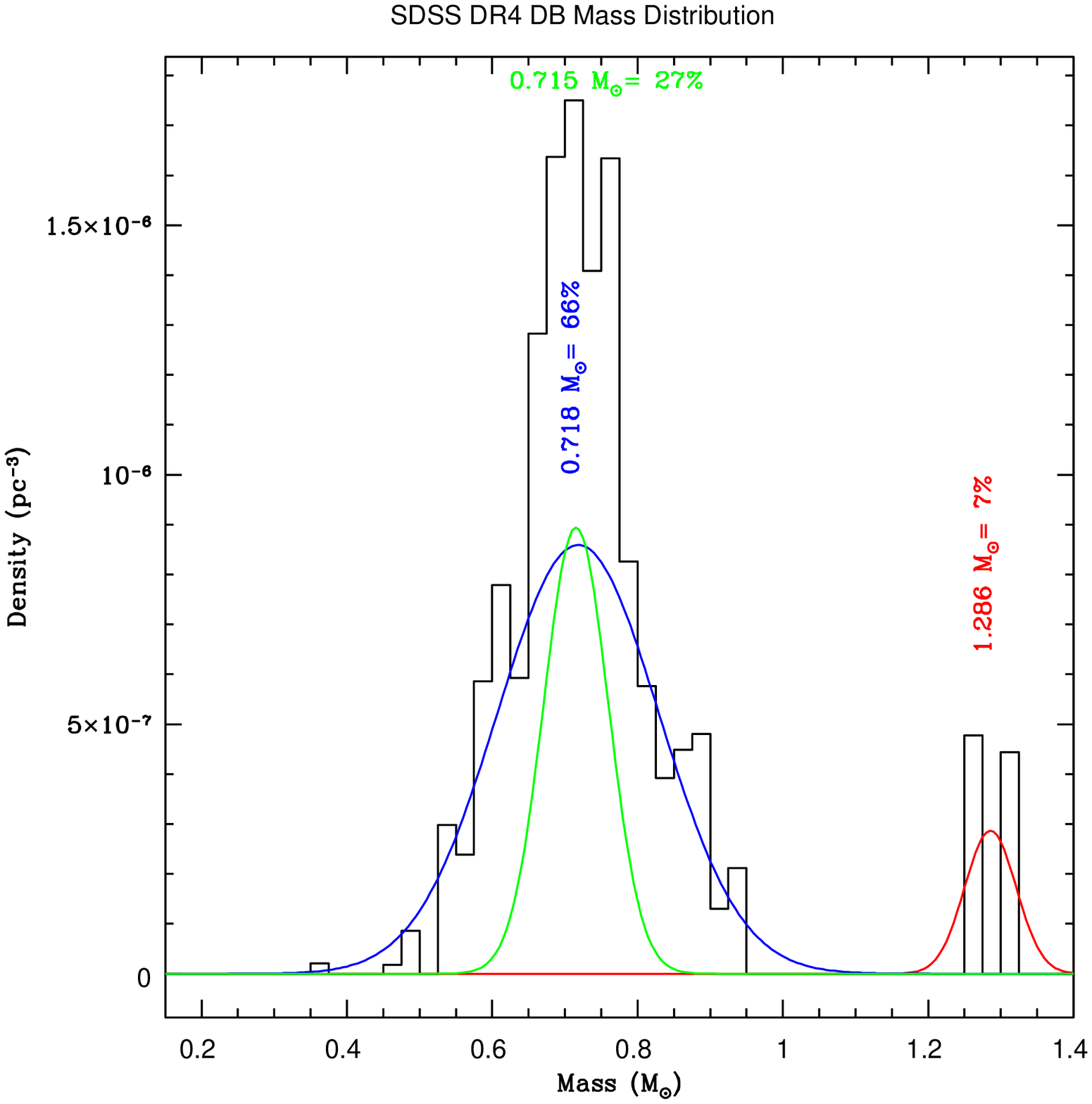}
      \caption{Histogram for the 150 DBs brighter than g=19, 
and hotter than
$T_{\mathrm{eff}}=16\,000$~K in DR4,
corrected by volume.
         \label{histvdb}
              }
   \end{figure}

Figures~\ref{histv}~and~\ref{histvdb} show the resulting corrected mass
distribution for our DA and DB sample, respectively. Figure~\ref{histv}
contains 1733 bright, non-cool DAs, i. e., those with
$T_{\mathrm{eff}}\geq 12\,000$~K
and $g\leq 19$.
We also list the corresponding five Gaussian fit parameters in 
Table~\ref{thistv1912}.
\begin{table}
\centering
\caption{Gaussian fits for the $T_{\mathrm{eff}}\geq 12\,000$~K
and $g\leq 19$ volume corrected DA mass histogram.
\label{thistv1912}}
\begin{tabular}{crccr}
\hline
i&$a_i$&$M_i ({\cal{M}}_\odot)$&$\sigma_i$&Fraction\\
\hline
1&$5.965\times 10^{-6}$&0.603&0.081&38\%\\
2&$1.203\times 10^{-5}$&0.571&0.034&32\%\\
3&$1.165\times 10^{-6}$&0.775&0.201&19\%\\
4&$1.455\times 10^{-6}$&1.175&0.076&9\%\\
5&$8.305\times 10^{-7}$&0.358&0.037&2\%\\
\hline
\end{tabular}
\end{table}
Figure~\ref{histvdb} contains 150 bright, non-cool DBs, i. e., those with
$T_{\mathrm{eff}}\geq16\,000$~K and $g\leq 19$.
The corresponding three Gaussian fits are listed in Table~\ref{thistvdb1916}.


Since the most massive white dwarf stars have
smaller luminosities because of their smaller radii, after applying
the $1/V_{\mathrm{max}}$ correction to the observed volume, we find
that around 20\% DAs are more massive than 0.8~${\cal{M}}_\odot$
in our bright and hot sample, of the same order as
that discovered by \citet{liebert} for the
PG sample.
The DB distribution is interesting, however, as it tends to
significantly higher masses than does the DA distribution! 
We found only two stars from our
sample with published atmospheric parameters
in the literature, 
with $\Delta T_{\mathrm{eff}}=510\pm 30$
and $\Delta \log g=0.12\pm 0.15$,
so we could not do a comparison
as \cite{scot} and \cite{scot06} did for the DA results.

\begin{table}
\centering
\caption{Gaussian fits for the volume corrected histogram of the 150 DBs brighter than g=19 and hotter than 
$T_{\mathrm{eff}}=16\,000$~K.
\label{thistvdb1916}}
\begin{tabular}{crccr}
\hline
i&$a_i$&$M_i ({\cal{M}}_\odot)$&$\sigma_i$&Fraction\\
\hline
1&$8.6\times 10^{-7}$ &0.718 & 0.111&66\%\\
2&$8.9\times 10^{-7}$ &0.715 & 0.045&27\%\\
3&$2.9\times 10^{-7}$ &1.286 & 0.031&7\%\\
\hline
\end{tabular}
\end{table}


\section{Extreme Mass Stars}
\citet{massive} published a catalog of all massive
white dwarf stars then known, with 112 stars
stars more massive than $0.8{\cal{M}}_\odot$.
The four stars with ${\cal{M}}\geq 1.3{\cal{M}}_\odot$
in their list are magnetic ones and therefore have large uncertainties
in their estimated masses. 
\citet{dahn} found one non-magnetic massive white dwarf,
LHS~4033,
with ${\cal{M}}\simeq 1.318-1.335{\cal{M}}_\odot$,
depending on the core composition.
Our oxygen-neon core mass for their derived $T_{\mathrm{eff}}=10\,900$~K and $\log g=9.46$
is ${\cal{M}}\simeq 1.30{\cal{M}}_\odot$.
We note that the models from the Montreal group
used to derive 
$T_{\mathrm{eff}}$ and $\log g$ in \citet{dahn}
show the same increase in mass with decreasing 
$T_{\mathrm{eff}}$ as our models and therefore we
do not take this mass determination as completely reliable due to the
objects relatively low temperature.
For GD~50 (WD J0348-0058), \citet{Dobbie} found
$T_{\mathrm{eff}}=41\,550\pm 720$~K and $\log g=9.15\pm 0.05$.
Our oxygen-neon core mass for their derived 
$T_{\mathrm{eff}}$ and $\log g$
is ${\cal{M}}\simeq 1.23\pm 0.02{\cal{M}}_\odot$,
very similar to the value reported by them for
C/O models.
They also show this massive star is consistent with its
formation and evolution as a single star, not the product of a merger.

\begin{figure}
\centering
   \includegraphics[width=\textwidth]{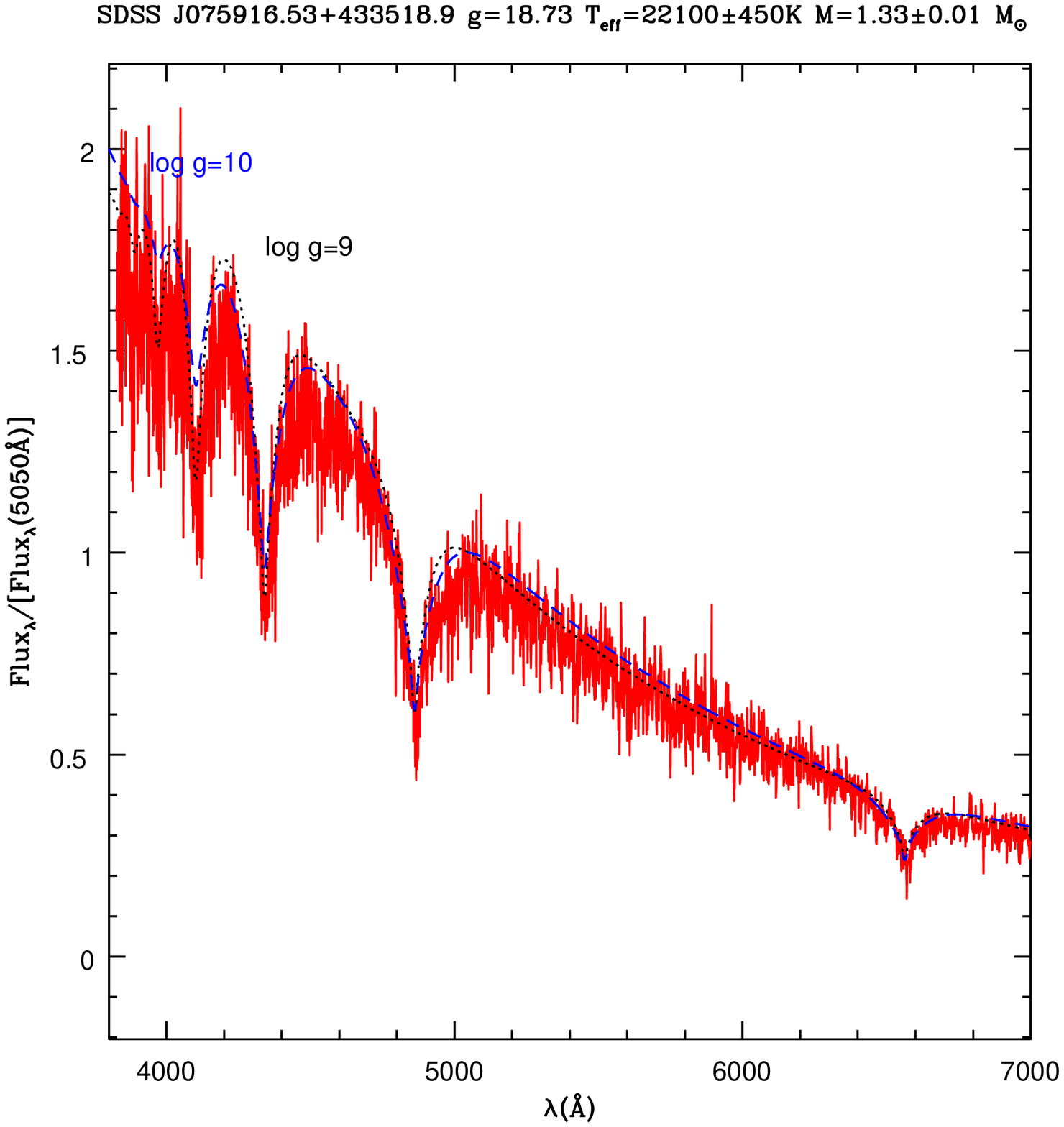}
      \caption{Spectrum of SDSS J075916.53+433518.9 with g=18.73,
$T_{\mathrm{eff}}=22\,100$~K and two models, with $\log g=9$ and 10.
The higher $\log g$ fits the H$\alpha$ line better, but the lower
$\log g$ fits the higher lines better, where the SNR is smaller.
This low SNR is typical for the stars closer to our upper cutoff of
g=19.
         \label{spspec}
              }
\end{figure}

From the 7167 pure DA white dwarf stars, we found 1611 (22\%)
with ${\cal{M}}> 0.8{\cal{M}}_\odot$. 
For the 2945 stars brighter than g=19
we found 760 (26\%) with ${\cal{M}}> 0.8{\cal{M}}_\odot$,
but for the 1733 stars brighter than g=19 and
hotter than $T_{\mathrm{eff}}=12000$~K, we find only 105 stars (6\%)
with ${\cal{M}}> 0.8{\cal{M}}_\odot$.
The most massive star in our
hot and bright sample is
SDSS J075916.53+433518.9, whose spectrum
spSpec-51883-0436-045 is shown in Fig.~\ref{spspec},
with g=18.73,
$T_{\mathrm{eff}}=22100\pm 450$~K,
$\log~g= 9.62\pm 0.07$,
${\cal{M}}=1.33\pm 0.01 {\cal{M}}_\odot$, and estimated distance of 
$d=104\pm 4$~pc.
We caution that the evolutionary models used to estimate the radii,
and therefore the masses, in our analysis do not include post-newtonian
corrections, important for masses above  ${\cal{M}}\simeq 1.30{\cal{M}}_\odot$
\citep{Chandra}.
For the stars brighter than g=19, we find 21 others
with masses larger than ${\cal{M}}=1.3~{\cal{M}}_\odot$, 
all below $T_{\mathrm{eff}}=9\,000$~K. We deem 
the mass determinations for stars cooler than $T_{\mathrm{eff}}\simeq 12\,000$~K unreliable.
In Table \ref{high12}, we list the DAs with $1.2~{\cal{M}}_\odot<{\cal{M}}<1.3~{\cal{M}}_\odot$
hotter than $T_{\mathrm{eff}}=12\,000$~K.

\begin{figure}
\centering
   \includegraphics[width=\textwidth]{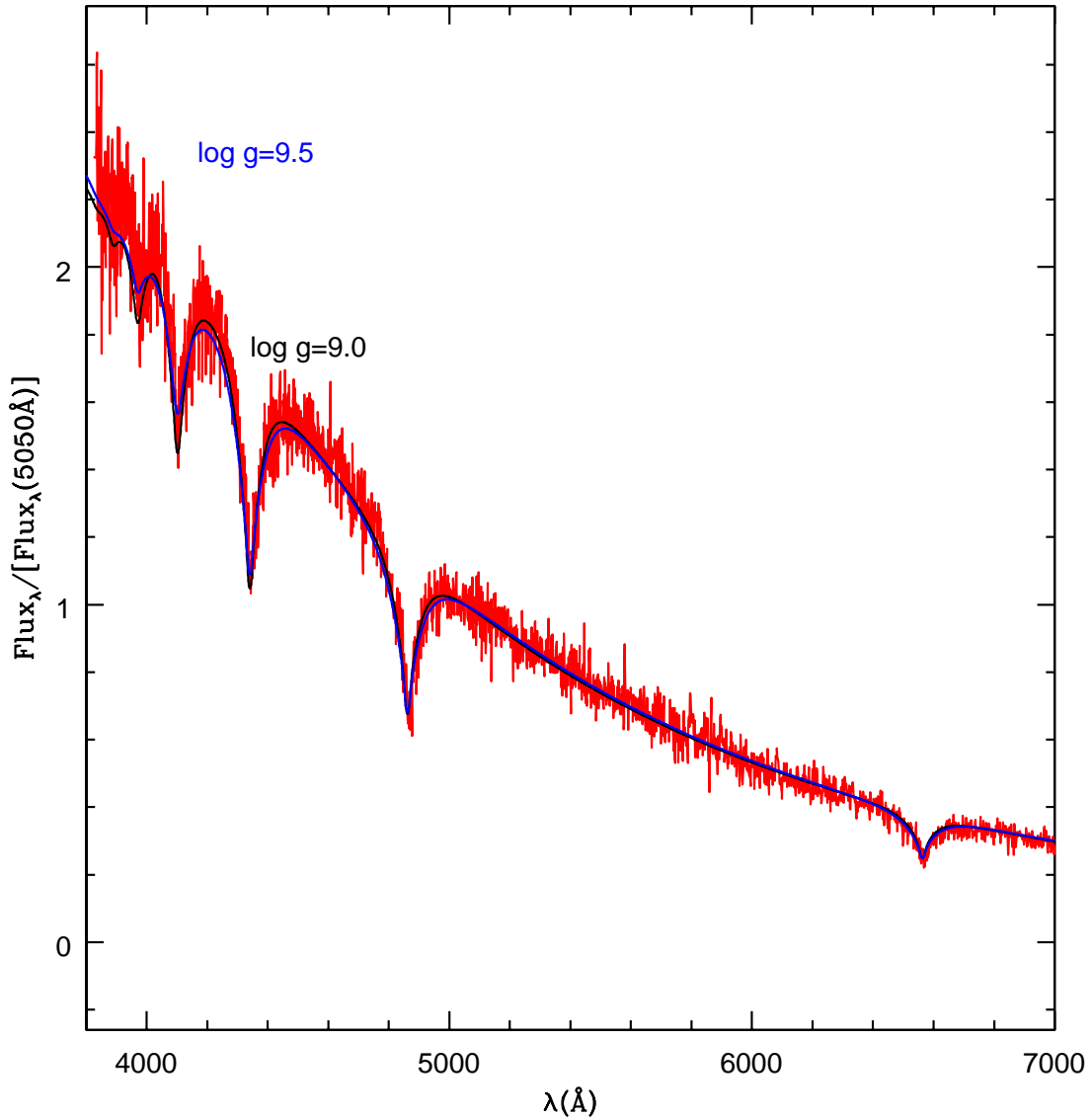}
      \caption{Spectrum of SDSS J094655.94+600623.4 with g=17.99,
$T_{\mathrm{eff}}=28\,100$~K and two models, with $\log g=9$ and 9.5.
This spectrum is typical of the the SNR achieved for the 1003 DAs and  59 DBs brighter than
g=18 in our sample.
         \label{spspec1}
              }
\end{figure}

\begin{table}
\begin{minipage}{126cm}
\begin{tabular}{lccccrccccr}
\hline
Spectra-M-P-F&Name&g&$M_g$&$T_{\mathrm{eff}}$&$\sigma_T$&$\log g$&$\sigma_g$&${\cal{M}}$&$\sigma_M$&d\\
             &    & &     &         (K)      &    (K)   &        &          &$({\cal{M}}_\odot)$&$({\cal{M}}_\odot)$&(pc)\\
\hline
spSpec-51691-0342-639&SDSS J155238.21+003910.3&18.44&12.23&15924&387&9.280&0.050&1.262&0.010&97\\
spSpec-51915-0453-540&SDSS J094655.94+600623.4&17.99&10.87&28125&220&9.370&0.040&1.287&0.010&123\\
spSpec-52374-0853-198&SDSS J133420.97+041751.1&18.52&12.34&17549&422&9.150&0.060&1.223&0.020&125\\
spSpec-52703-1165-306&SDSS J150409.88+513729.1&18.84&10.28&79873&8228&9.050&0.390&1.204&0.100&468\\
spSpec-52751-1221-177&SDSS J110735.32+085924.5&18.42&12.23&18715&327&9.140&0.060&1.219&0.020&128\\
spSpec-52872-1402-145&SDSS J154305.67+343223.6&18.33&10.85&30472&313&9.300&0.070&1.269&0.010&168\\
\hline
\end{tabular}
\caption{DA stars with masses above $1.2~{\cal{M}}_\odot$ and below  $1.3~{\cal{M}}_\odot$ derived from the SDSS spectra,
with $T_{\mathrm{eff}}\geq 12\,000$~K. 
}
\label{high12}
\end{minipage}
\end{table}
The spectrum for the brighter g=17.99
SDSS J094655.94+600623.4 is shown in Figure~\ref{spspec1}.
Because our analysis uses relatively low SNR spectra and 
gravity effects dominate mainly below 3800\AA, where we have no data,
our conclusion is that we must undertake a study in the violet
or ultraviolet to measure the masses more accurately.
An extensive study of gravitational redshift would
also be critical.

For the 507 single DBs 
we find 30 DBs with $\log g>9$.  Most of our massive DBs 
are cooler than $T_{\mathrm{eff}} \simeq 16\,000$~K, or fainter than g=19,
except for SDSS J213103.39+105956.1  with g=18.80,
$T_{\mathrm{eff}}= 16476\pm 382$, and $\log g=9.64\pm0.21$,
corresponding to a mass ${\cal{M}}=1.33\pm 0.04 {\cal{M}}_\odot$,
and for SDSS J224027.11-005945.5 with g=18.82,
$T_{\mathrm{eff}}= 17260\pm 402$, and $\log g=9.31\pm0.20$,
corresponding to a mass ${\cal{M}}=1.25\pm 0.06 {\cal{M}}_\odot$.

\begin{figure}
   \centering
   \includegraphics[width=\textwidth]{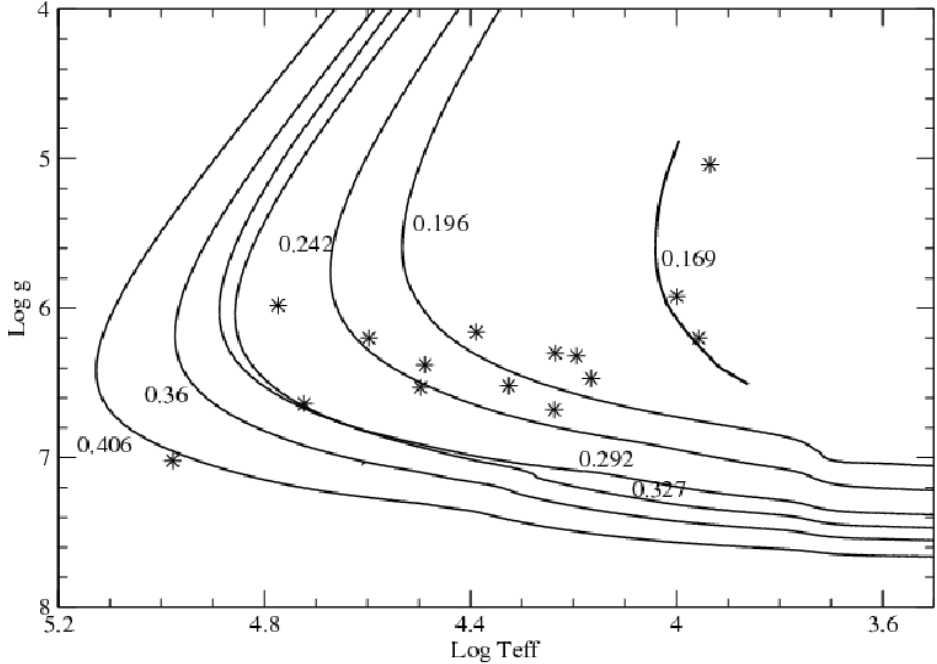}
      \caption{Evolutionary tracks for He white dwarf stars
calculated by \citet{Althaus01} and the location of the
lowest mass stars in our sample. Most of these stars are
below g=19 and therefore have noisy spectra.
         \label{lowmass}
              }
   \end{figure}

The low mass stars present in our sample are consistent with He core evolution
models calculated by \citet{Althaus01}, and displayed in Fig.\ref{lowmass}.
It is important to stress that these stars should be studied with more accurate
spectra and model atmospheres, as they are possible progenitors of SN~Ia
if they accrete mass from companions.

\section{Conclusions}

\begin{figure}
\centering
   \includegraphics[width=\textwidth]{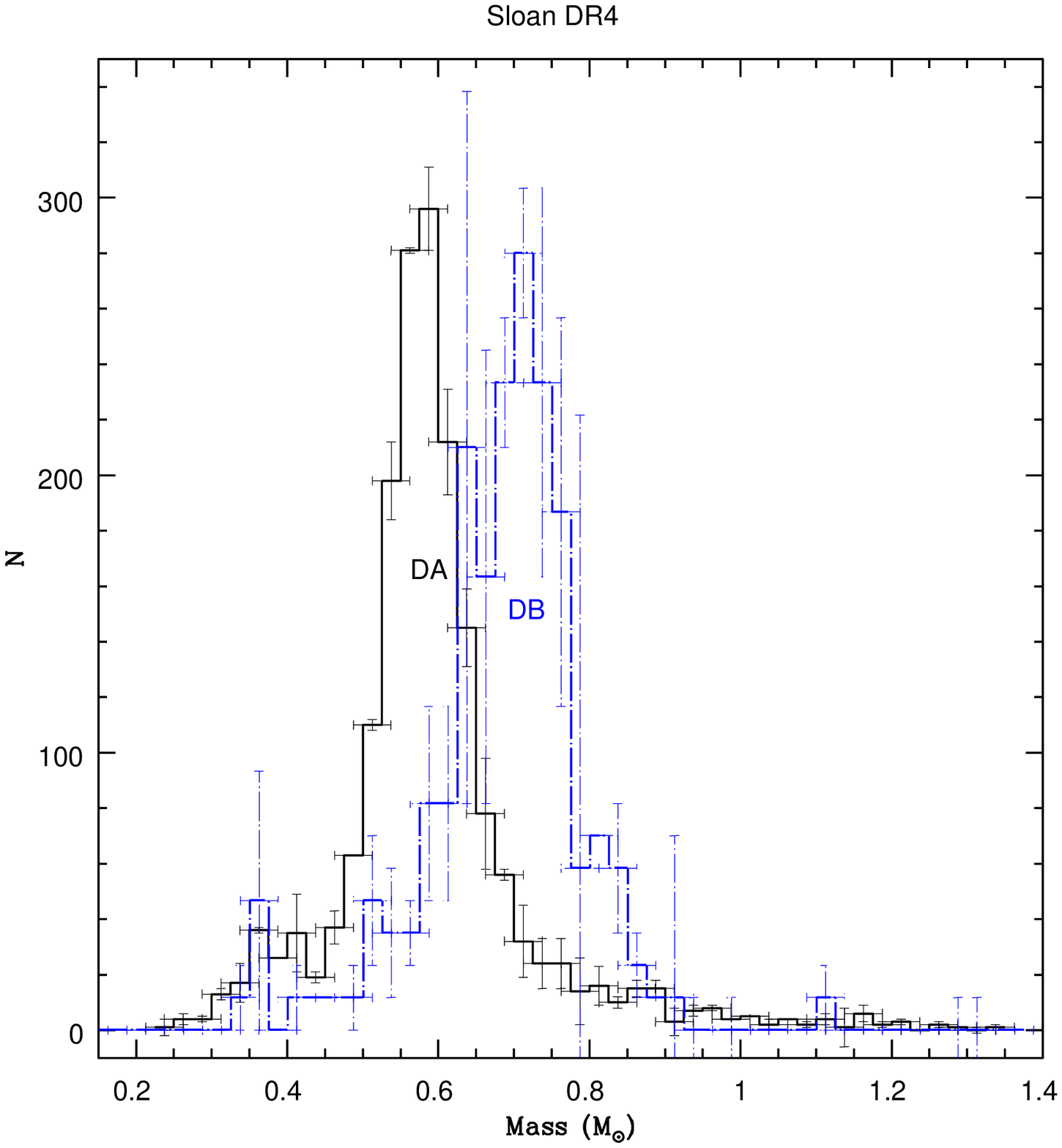}
      \caption{DA and DB histograms 
for comparison. 
The DB histogram has been re-normalized to the DA maximum for display
purposes.
         \label{histe}
              }
\end{figure}
\begin{figure}
\centering
   \includegraphics{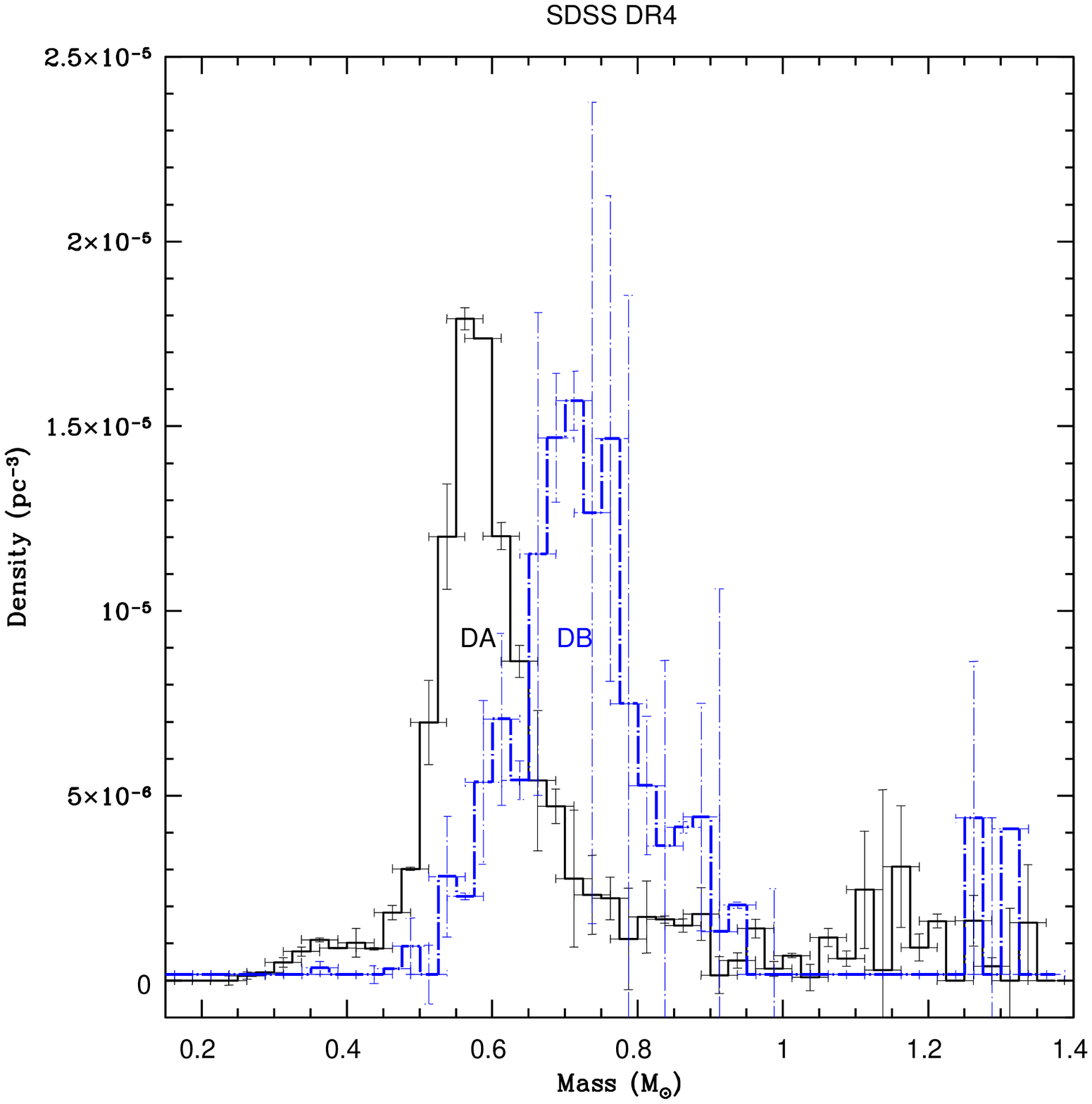}
      \caption{DA and DB histograms corrected by 
observed volume for comparison. 
The DB histogram has been re-normalized to the DA maximum for display.
         \label{histvdadb}
              }
\end{figure}
Our investigations into the mass distribution of the SDSS DR4 white dwarf
sample from \citet{scot06} revealed several items.  First, all groups are
seeing nearly identical increases in mean DA at lower
temperatures (less than $\approx 12\,000$K for DAs and $16\,000$K for DBs).
Either this is truly going on in the white dwarf stars, or
there is missing or incorrect physics in everyone's models. We propose the
treatment of neutral particles as the most likely explanation.
We suspect the atmospheric models should be improved with a detailed
inclusion of the line broadening by neutral particles, since
the increase in apparent mass for both DAs and DBs occur
at temperatures when recombination becomes important.

Secondly, we find a significant difference between the DA and DB mass
distributions, with the DB distribution significantly more weighted to
massive stars.
Figs.~\ref{histe} 
and \ref{histvdadb} 
show the combined number and volume-corrected DA and DB histograms.
The DB histograms have been re-normalized to the DA 
maximum for display purposes. 
Our results contradict nearly all previous work which
show the mean DA and DB masses to be similar 
\citep[with the exception of][]{koester2001}.
We note that the
previous efforts, though, were based on histograms for DBs with less than
50 stars, and our DB histogram has 150 stars.  
However, we still need to explore our DB
models and fits in more detail to verify the validity of this novel result.
Specifically, we find $\langle {\cal{M}}\rangle_\mathrm{DB}=0.711\pm 0.009 {\cal{M}}_\odot$,
higher than 
$\langle {\cal{M}}\rangle_\mathrm{DA}=0.593\pm 0.016 {\cal{M}}_\odot$ for the
1733 DAs brighter than g=19, and hotter than $T_{\mathrm{eff}}=12000$~K.
This is a significant new result and must be investigated further.

We have also detected a large number of massive DA white dwarf stars:
760 with ${\cal{M}}> 0.8{\cal{M}}_\odot$ brighter than g=19
and 105 both brighter than g=19 and
hotter than $T_{\mathrm{eff}}=12000$~K.
We report the highest $\log g$ white dwarf ever detected.

\section*{Acknowledgments}

We thank Dr. Daniel Eisenstein for some helpful suggestions, and his {\it autofit} code,
and the referee, Dr. Martin Barstow, for very detailed and useful comments that improved
the presentation of the paper.

\bsp

\label{lastpage}

\end{document}